\documentclass[aps,prl,twocolumn,superscriptaddress]{revtex4}
\newcommand{\figwidth}{3.0in} 

\usepackage{amssymb}
\usepackage{graphicx}

\newcommand{\ket}{\rangle}
\newcommand{\bra}{\langle}

\newcommand{\Yj}{Y-junction}
\newcommand{\Yjs}{Y-junctions}
\newcommand{\Ydelj}{Y$_{\Delta}$-junction}
\newcommand{\Ydel}{Y$_{\Delta}$}
\newcommand{\half}{\frac{1}{2}}

\begin{document}
\bibliographystyle{apsrev}

\title{Density matrix renormalization group algorithms for \Yjs } \author{
  Haihui Guo} \author{ Steven R.\ White} \affiliation{ Department of Physics
  and Astronomy, University of California, Irvine, CA 92697 } \date{\today}
\begin{abstract}
  \noindent
Systems of \Yjs \ are interesting both from a fundamental viewpoint and
because of their potential use in nanoscale devices. These systems can be studied
numerically with the density matrix renormalization group(DMRG), but existing algorithms
are inefficient.  Here, we introduce a much more efficient DMRG algorithm 
for \Yj \ systems. As an example of the use of this method, we study $S=\half$ 
bound states in Heisenberg $S=1$ junctions with two geometries, one where the junction
consists of a single site, and the other where it consists of a triangle of three sites.

\end{abstract}
\pacs{PACS Numbers: }
\maketitle
Junctions are essential ingredients of existing and future electronic and
spintronic devices. As progress towards smaller and smaller devices
continues, eventually we will reach the atomic scale. Perhaps the smallest
conceivable junctions are composed of intersecting spin or electron chains, 
with each site representing a single atom. Isolated chains, modeled by
Heisenberg, Hubbard, and related Hamiltonians, have been extensively studied.
Yet surprisingly little work has been done on the corresponding junctions of
three or more half-chains, or legs.  There has been recent work on junctions of quantum
wires. For example, Oshikawa, {\it et al}\cite{COA03,OCA06} studied the
transport properties of a junction composed of three quantum wires enclosing
a magnetic flux, modeling the chains as spinless Tomonaga-Luttinger liquids,
and finding a rich phase diagram.  Quantum Monte Carlo on junctions of
fermion chains are subject to the minus sign problem, so little numerical
work has been done. We are not aware of any studies which have focused on junctions of spin
chains.
 
Our present work has two main aspects. First, we develop a new density matrix
renormalization group (DMRG) algorithm for studying junctions. Ordinary
DMRG\cite{dmrg} is much less efficient for junctions compared to chains with
open boundaries; our new method requires only slightly more computational
effort. Second, as an application of this method we consider a key feature
of a few of the simplest junctions, namely the presence of
$S=\half$ ``spinon'' bound states in two types
of $S=1$ Heisenberg Y junctions. 
These bound states are the analogues of the
spin-$\half$ end states found in open $S=1$ chains\cite{WH93,GGLKM91}. 
They
exist at any $S=1$ junction with an odd number of legs, {\it regardless of
  the local interactions at the junction}. Magnons coming to the junction can
interact with the bound state via spin-flip scattering, leading to
potentially interesting dynamical phenomena. We call these spinon bound states because a
spin-$\half$ degree of freedom is bound to the junction, but it is important to note
that they are features of the ground state. A separate issue, which we do
not address here, is whether the
lowest excited state is below the Haldane gap, representing, say, a magnon bound to the junction.
For example, a single $S=\half$ impurity in a $S=1$ chain has such a bound state below the
gap only for sufficiently weak coupling to the impurity\cite{sorensen}.
 
\begin{figure}[tb]
  \includegraphics*[width=\figwidth]{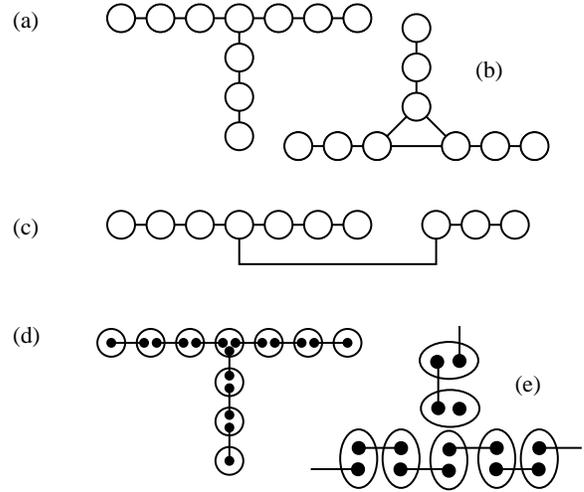}
  \caption{ Schematic view for the geometric configuration and algorithm for
    \Yjs.The geometry of the \Yj is shown in (a), where the circles represent
    sites and the solid lines represent connections. The alternative geometry
    \Ydel \ junction is shown in (b). The original DMRG method is shown in (c),
    where one cuts the link between two sites and relinks it to the center of
    the remaining chain, making it topologically equivalent to a \Yj.
    Verstraete and Cirac's tensor representation is shown in (d), where each open circle
represents a tensor and each solid dot represents a tensor index. An
    Affleck-Kenedey-Lieb-Tasaki valence bond picture of the \Yj \ is shown in (e), where
each solid dot represents a $S=\half$, the line segments represent valence bonds,
and where each open circle
represents a projection operator to the $S=1$ subspace for a site.  }
  \label{figone}
\end{figure}
 
We consider two types of three-leg junctions, which we will label Y and
\Ydel, where we think of the $\Delta$ as a triangle. These are shown in
Fig. \ref{figone}(a) and (b). DMRG is extremely efficient for one dimensional
systems with short range interactions. It is standard to treat
two-dimensional strips by mapping the strip into a chain with interactions
over a distance of several lattice spacings. Similarly, a Y junction can be
treated as a chain with one long range interaction, shown in Fig. \ref{figone}(c).
The long-range interaction is equivalent to periodic
boundary conditions in terms of the effort necessary in DMRG: a typical block
of sites connects to the rest of the system by two links in both cases. If
$m$ states per block are required for a given accuracy for a single chain,
one expects roughly $m^2$ states to be needed for the Y junction (or a periodic chain).
This changes the computational effort from $O(N m^3)$ to $O(N
m^6)$, where $N$ is the number of sites.

As discussed below, our new algorithm scales as $O(N m^3 + m^4)$. Typically $m$ and $N$
are similar in size (say $\sim 10^2$), 
so the cost is close to that of a single chain, and a large improvement
over the method with a long-range interaction.
Our algorithm is somewhat related to Otsuka's early DMRG treatment of the XXZ model on a
Bethe lattice\cite{Otsuka}. 
It is closely related to Shi, Duan, and Vidal's very recent
treatment\cite{Tree} of general tree-tensor networks using Vidal's
time-evolving block decimation method, which is closely related to DMRG. One
can also view it as perhaps the simplest step towards an implementation of
Verstraete and Cirac's
much more complicated tensor approach\cite{mps,PEPS}, which
holds great promise for true two dimensional systems. 

Consider the matrix product representation for the wavefunction in
standard DMRG\cite{OR95}
\begin{equation}
  |\Psi \rangle = A[s_1]A[s_2]A[s_3]...A[s_n]|s_1 s_2 s_3...s_n\rangle
\end{equation}
where $s_i$ labels the states of site $i$, and $A[s_i]$ is a matrix except for
the first and last sites, where it is a vector.
One can view each matrix as a two-dimensional tensor, with indices connecting
to the left and right. In Verstraete and Cirac's approach for a square
lattice, one generalizes to four-dimensional tensors, with indices connecting
to each nearest neighbor. In our approach, only one tensor is needed, at the
junction site, and it has only three indices. All sites on the legs are
ordinary matrices, as shown in Fig. \ref{figone}(d). Unlike in the more
general tensor approach, the states can be easily described in terms of an
orthogonal basis.


\begin{figure}[tb]
  \includegraphics*[width=\figwidth]{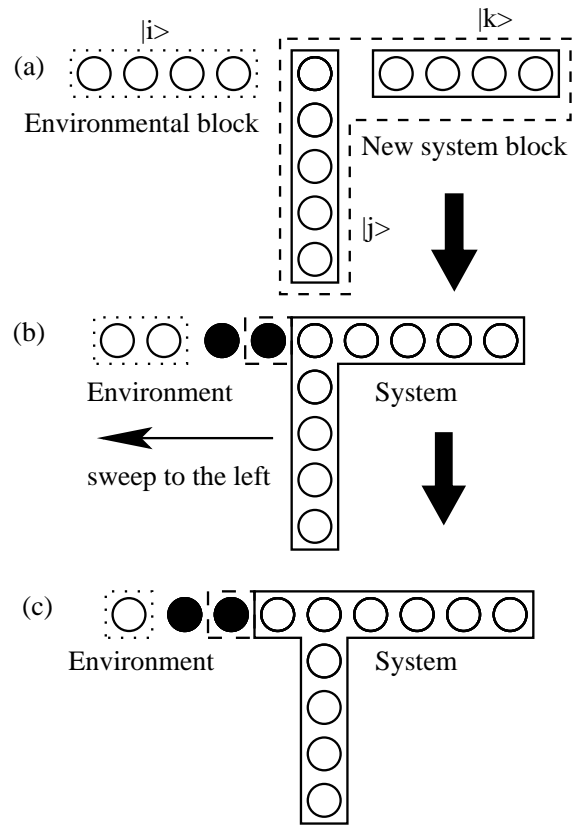}
  \caption{ Configurations illustrating the new DMRG algorithm. The procedure starts
    from the center (a), where two legs are combined to form a new system
    block. The sweep next progress down a single leg, as in (b), reaches
    the end of the leg, (c), and comes back to the center.  }
  \label{figtwo}
\end{figure}

The key to our approach is a new type of step for the junction. In this ``junction step''
the system is divided into three blocks instead of two (one for each leg), and the calculation
time is higher, $O(m^4)$, where $m$ is the number of states kept per block, instead of
$O(m^3)$. The rest of a sweep consists of steps involving two blocks, where one of the
blocks contains two legs plus part of the third. 
The sweep moves out to the end of one leg, and then back to
the center, after which the junction step is repeated, and then the sweep goes
out another leg, etc., until all legs have been treated. 

We now describe this method in more detail.
The system is built up in a warmup sweep. An ordinary DMRG chain
initialization sweep is first carried out on one leg of the \Yj. For the
environment block one can use the next several sites of the leg, or
artifically reflect the leg--the algorithm is not very sensitive to the
warmup, and we keep only a small number of states $m$, storing the related
operator matrices and transformation matrices. One continues until the entire
leg is incorporated into the block. At the last few steps the environment
may or may not be in the form of a junction---it does not make much
difference. This block is then duplicated twice to form the other two
legs.

At this point, the finite-system sweeps begin. The first step is the junction step.
For the \Yj, where the junction consists of a single site, that site is incorporated
into one leg.
The wavefunction can be written as $\psi_{ijk}$, where $i$, $j$, and $k$ refer to
the three legs/blocks.  Applying the
operators appearing in the Hamiltonian $H$ to $\psi$ requires a calculation time of $O(m)^4$,
in contrast to the $O(m)^3$ time on a chain. For example, to apply the term
$S_{ii'}^z S_{jj'}^z$ connecting the first two legs, one first multiplies
the matrix $S_{jj'}^z$ into $\psi_{i'j'k}$, storing the result, and then
multiplies that result by $S_{ii'}^z$. Several such multiplications by $H$
are performed to make a few Lanczos or Davidson steps to find an approximate ground state
$\psi$.

After finding $\psi$ we want to combine two legs (say $j$ and $k$) into a single effective block,
as shown in Fig. \ref{figtwo}(a). In
ordinary DMRG, we find the reduced density matrix for the system block, and
diagonalize it, which in this case would result in a calculation time of $O(m^6)$.
Instead, we perform a singular value decomposition (SVD) on $\psi$, treating
$j$ and $k$ together as a single combined index, and $i$ as the other index
of the matrix. The SVD has a calculation time of only $O(m^4)$, but is equivalent to the
density matrix diagonalization. (Note, however, that the SVD is not as
easily generalized to having more than one state targeted.)  One obtains transformation
matrices $O_{\alpha ,jk}$, where $\alpha$ are the new states of the combine
blocks. The operators for the combined block $A'$ are obtained as $OAO^T$;
this can also be performed in $O(m^4)$ time. For example, to transform
$S^+_{jj'}S^-_{kk'}$, one first forms $X_{\alpha ,j'k}=O_{\alpha
  ,jk}S^+_{jj'}$, then $Y_{\alpha' ,j'k}=S^-_{kk'}O_{\alpha' ,j'k'}$, and
then contracts $XY$ over indices $j'k$.

Next, as shown in Fig. \ref{figtwo}(b), we sweep down leg $i$. This sweep is
an ordinary DMRG sweep, with calculation time $O(m^3)$ per site; the fact
that the system block now includes two legs does not affect calculations in
each step. We continue moving to the edge of the leg, Fig. \ref{figtwo}(c),
and then turn around and come back to the center. The junction step is
performed and then the sweep moves out one of the other legs. The three
legs need not be identical; the sweep treats each independently. In each
full sweep, the center step is performed three times. Alternatively, if
the Hamiltonian and the desired state are symmetric with respect to the
interchange of legs, just before the junction step one can replace each of
the legs with the most recently updated leg/block.

\begin{figure}[tb]
  \includegraphics*[width=\figwidth]{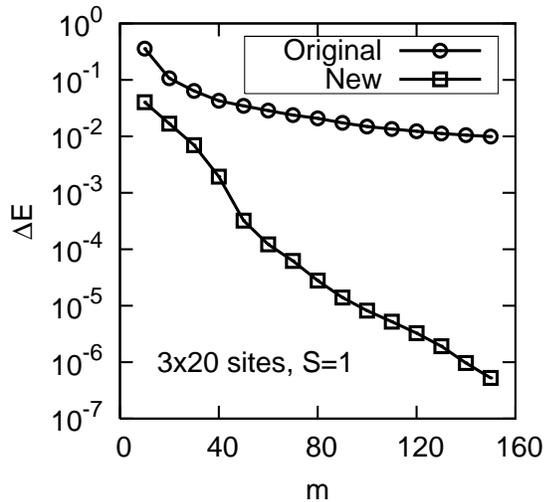}
  \caption{ Error in the energy as a function of the number of states kept
    $m$ for the new and the standard DMRG algorithms. The reference energy
    $E$ here is the result of standard algorithm by keeping $3000$ states.  }
  \label{figthree}
\end{figure}

For an $3\times N$ system, the total cpu time is $O(m^4+N m^3)$. In
comparison, we consider the standard DMRG calculation with one long range
interaction. In Fig. \ref{figthree} we show the error in the energy as a
function of the number of states kept for the new versus the standard
algorithm; the new algorithm requires vastly fewer states. If we assume the
long range interaction mandates $O(m^2)$ states be kept for sufficient
accuracy (for large $N$), the standard DMRG calculation would scale as $O(Nm^6)$.

We now use the new algorithm to study $S=\half$ bound states in $S=1$
Heisenberg junctions. A simple consideration of quantum numbers makes the
presence of the bound states clear. An $S=1$ open chain has $S=\half$ states
on the ends, and a finite correlation length $\xi=6.03$\cite{WH93}. The
ends of the legs in a large junction system will also have $S=\half$
states; the finite correlation length does not allow the junction to
influence the ends. With an odd number of legs, a half-integer-spin state
must form at the junction to make the total spin an integer. For a junction
formed from very weakly coupled legs, the bound state comes from coupling
three effective $S=\half$ states at the junction. For more strongly coupled
legs one can think in terms of the Affleck-Kennedy-Lieb-Tasaki (AKLT) picture\cite{AKLT}
where each $S=1$ is composed of two $S=\half$'s, with one valence bond
singlet connecting two $S=\half$'s on each near-neighbor link; see Fig. 1(e). At the
junction, there must be one $S=\half$ left over after the singlets are
formed.

\begin{figure}[tb]
  \includegraphics*[width=\figwidth]{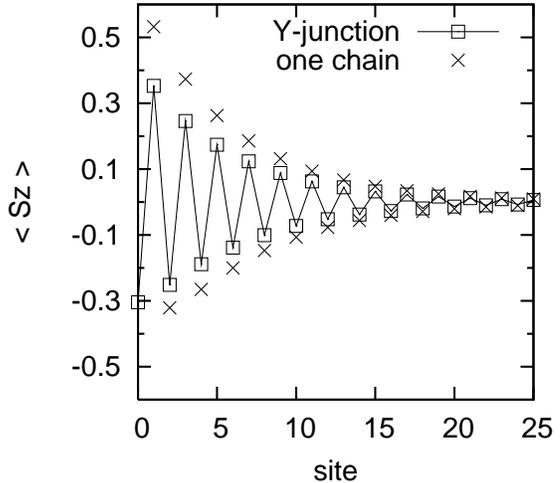}
  \caption{ Results for $\bra Sz(l) \ket$ for a \Yj \ as a function of site
    $l$ along a chain, compared with an open 1D chain. For the \Yj, the
    first site shown is the center junction site. Here, the total spin of the
    system is $S_z=\half$.}
  \label{figfour}
\end{figure}

In order to study the junction bound state, it is convenient to place a real
$S=\half$ spin on the end of each chain (far from the junction), which forms
a singlet with the effective $S=\half$, eliminating it as a low energy degree
of freedom. For the \Yj, the ground state is then a nondegenerate doublet
corresponding to total $S_z=\pm \half$. In Fig. \ref{figfour} we show $\bra
S_z(l) \ket$ as a function of site $l$ for a $3\times60$ \Yj. The result for the other
two legs is identical. We see a strong similarity with the results for the
end state of a single chain, except for a smaller magnitude of the
oscillations. The decay of the oscillations is consistent with an exponential
decay with the correlation length of a single
chain\cite{WH93}. Looking near the junction we find the initially
surprising result that the value of $|\bra S_z \ket|$ at the junction site
(0.30) is less than that on the first site of a leg (0.34). If one thinks
in terms of a mean-field/classical picture, with the spin fluctuations acting
like thermal fluctuations, one would expect the opposite: the junction site
is surrounded by three polarized spins, and would experience a larger
polarizing field than the adjacent sites which have only two neighbors.
However, if one utilizes the AKLT valence bond picture instead, one would
predict a smaller magnitude at the junction site: the extra unpaired
$S=\half$ is energetically favored to be on an adjacent site, with one
missing valence bond, as shown in Fig. \ref{figone}(e). An unpaired spin on
the junction site occurs from fluctuations of the unpaired $S=\half$ from leg to leg. 

\begin{figure}[t]
  \includegraphics*[width=\figwidth]{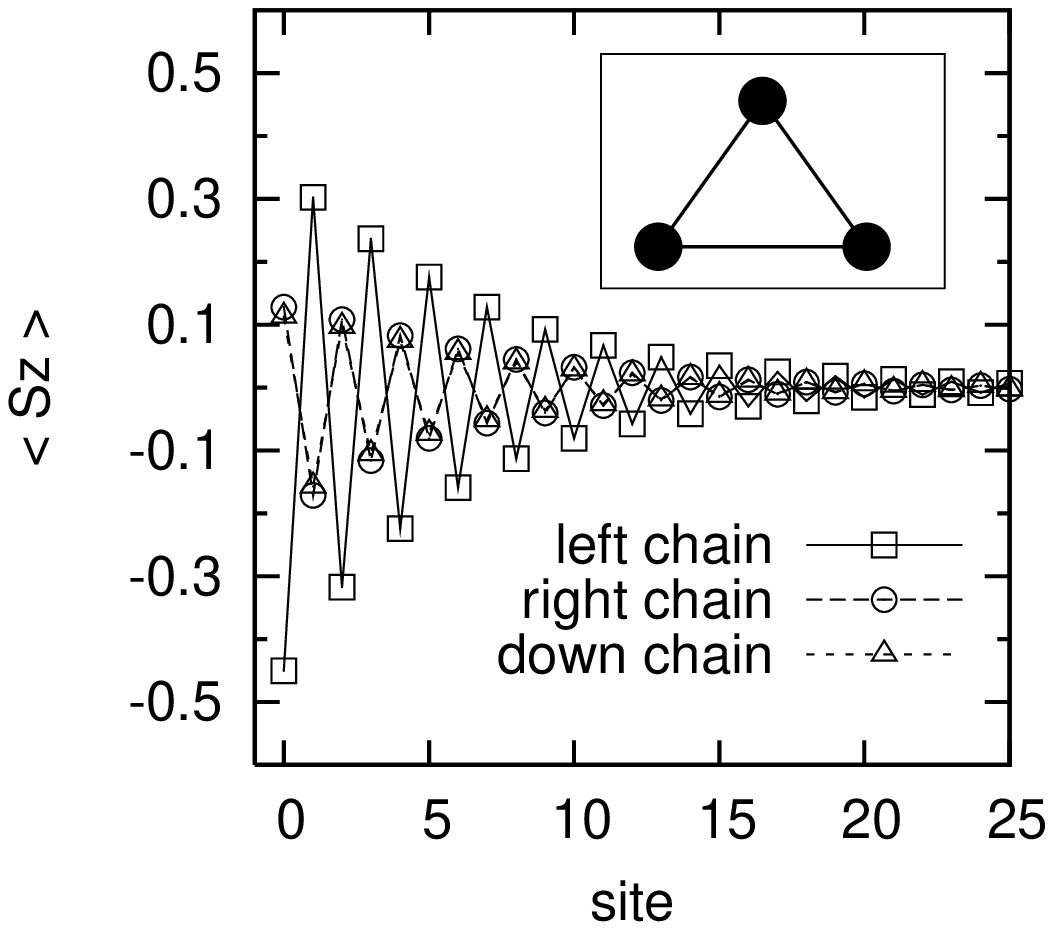}
  \caption{ Results for $\bra Sz(l)\ket$ for a $3\times 60$ \Ydelj, 
where site 0 is part of the junction triangle. 
 }
  \label{figfive}
\end{figure}

Now consider the other geometric configuration, the \Ydelj.
Fig. \ref{figfive} shows $\bra S_z(l) \ket $ for a \Ydelj  with size
$3\times 60$. One sees the presence of a bound state at the junction, as for
the \Yj. However, the symmetry under interchange of legs is broken by an
exact ground state degeneracy.

To understand this degeneracy, consider a three-site $S=\half$ Heisenberg triangle. The
$S=\half$ triangle is clearly relevant in the limit that the
exchange couplings connecting the three junction sites to each other is very
small. In that case the $S=\half$'s are the end states of each leg. In the
opposite limit, where the exchange between the $S=1$'s in the junction is
large, the ground state of the $S=1$ triangle is nondegenerate (with a gap of precisely $J$), 
but then one
expects a $S=\half$ end state on each leg starting at the site adjacent to
the junction, and one would expect these three $S=\half$'s to be weakly coupled
antiferromagnetically through the central triangle.

If we square the total spin operator for a generic Heisenberg triangle, we
obtain
\begin{equation}
  \label{eq:triangle}
  \frac{E}{J}=\frac{1}{2}(S_{\rm tot}(S_{\rm tot} +1)-S_1(S_1+1)-S_2(S_2+1)-S_3(S_3+1)) .
\end{equation}
Here $S_1 = S_2 = S_3 = \frac{1}{2}$, and $S_{\rm tot}$ can be either
$\frac{1}{2}$ or $\frac{3}{2}$. The ground states have 
energy $-\frac{3}{4} J$ and consist of two 
degenerate $S_{\rm tot}=\frac{1}{2}$
doublets. If one regards the triangle as a periodic chain, then the degeneracy
corresponds to total momenta around the triangle of $k=\pm 2 \pi/3$. 
The $S_{\rm tot}=\frac{3}{2}$ multiplet has $k=0$ and energy $\frac{3}{4} J$. 
The \Ydelj \ has the
same degeneracy as the ground states of the triangle; 
the numerical calculation converges to a arbitrary linear
combination of these states. We have confirmed the degeneracy behaves as
expected by targetting several states on a relatively small \Ydelj \ system.

In conclusion, we have presented a relatively simple DMRG algorithm for
junction systems, with greatly improved computational effort. We have used
this to study the $S=\half$ bound states of two types of Heisenberg $S=1$
\Yj.

We thank S. Chernyshev, F. Verstraete, and I. Affleck for helpful conversations. We
acknowledge the support of the NSF under grant DMR03-11843.

{}


\begin{thebibliography}{}

\bibitem{COA03} C.~Chamon, M.~Oshikawa, and I.~Affleck, \prl {\bf 91},
  206403(2003).
\bibitem{OCA06} M. Oshikawa, C. Chamon, and I. Affleck, J. Stat.
  Mech(2006)P02008.
\bibitem{dmrg} S.R.\ White, Phys.\ Rev.\ Lett.\ {\bf 69}, 2863 (1992); Phys.\
  Rev.\ B {\bf 48}, 10345 (1993).
\bibitem{WH93} S.R. White, D.A. Huse, \prb {\bf 48},3844(1993).
\bibitem{GGLKM91} S.H.\ Glarum, S.\ Geschwind, K.M.\ Lee, M.L.\ Kaplan, and
  J.\ Michel, \prl {\bf 67}, 1614(1991).
\bibitem{sorensen} Bound states (with $S=1$) have been found on the weak links
of $S=1$ chains; see E.\ Sorensen and I. Affleck \prb {\bf 51}, 16115(1995).
\bibitem{Otsuka} H.\ Otsuka, \prb {\bf 53}, 14004(1996).
\bibitem{Tree} Y.\ Shi, L.\ Duan and G.\ Vidal, quant-ph/051107
\bibitem{mps} F.\ Verstraete, D.\ Porras, and J.I.\ Cirac, Phys.  Rev.  Lett.
  {\bf 93}, 227205 (2004).
\bibitem{PEPS} F.\ Verstraete and J.I.\ Cirac, cond-mat/0407066
\bibitem{OR95} S.\ Ostlund and S.\ Rommer, \prl {\bf 75}, 3537(1995); S.\
  Rommer and S.\ Ostlund, \prb {\bf 55}, 2164(1997).
\bibitem{AKLT} I. Affleck, T.Kennedy, E.H. Lieb and H. Tasaki, \prl {\bf 59},
799(1987).

\end{thebibliography}
\end{document}